\documentclass{ws-procs9x6}
\newcommand{\<}{\langle}
\renewcommand{\>}{\rangle}
\def\lsim{\raise0.3ex\hbox{$<$\kern-0.75em\raise-1.1ex\hbox{$\sim$}}}
\def\gsim{\raise0.3ex\hbox{$>$\kern-0.75em\raise-1.1ex\hbox{$\sim$}}}

\begin{document}

\title{On the renormalization of the Polyakov loop
\footnote{\uppercase{S}upported in part by
    the \uppercase{DFG} under grant \uppercase{FOR} 339/1-2. }}

\author{F. Zantow}

\address{Fakult\"at f\"ur Physik, Universit\"at Bielefeld, D-33615 Bielefeld,
  Germany\\
E-mail: zantow@physik.uni-bielefeld.de}


\maketitle

\abstracts{We discuss a non-perturbative renormalization
  of $n$-point Polyakov loop correlation functions by
  explicitly introducing a renormalization constant for the Polyakov loop
  operator on a lattice deduced from the short distance properties of $2$-pint
  correlators. We calculate this constant for the $SU(3)$ gauge
  theory.}
\section{Need of renormalized Polyakov loop}
Universal
properties of the confinement-deconfinement phase transition in QCD
are often discussed in the quenched
approximation ($\sim SU(N_c\equiv3)$) where the Polyakov loop $L$ is treated
as an order parameter. In a lattice regularized version, $L({\bf
  x})=\prod_{x_0=1}^{N_\tau}U(x_0,{\bf x})$, where $U(x_0,{\bf x})\in
SU(N_c)$ denotes the gauge link variable on a lattice
of size $N_\sigma^3\times N_\tau$. The
expectation value of the traced Polyakov loop is said to be
related to the free energy\cite{McLerran} of a single
test-quark put into a
gluonic heat bath ($F_q$) and that of the heat bath alone ($F_0$) via:
\begin{eqnarray}
\<L\>&\equiv&\<\frac{1}{N_\sigma^3}\sum\limits_{{\bf x}}\tilde{\mbox{Tr}}L({\bf x})
\>\;\simeq\;\frac{Z_q}{Z_0}\;=\;e^{-\beta(F_q-F_0)}.
\qquad\quad(\tilde{\mbox{Tr}}1\!\!1\equiv1)
\label{Pev}
\end{eqnarray}
$Z_q$ and $Z_0$ denote the partition functions of the system with
and without test-quark\footnote{$Z_q$ defined through (\ref{Pev}) vanishes on
  any finite lattice because of the $Z_N$ symmetry of $L$. For a more formal
  definition of $\<L\>$ and
  $Z_q$ see Ref. 5.} at inverse temperature $\beta=1/T$. However, it is
well-known that $\<L\>$ contains (UV-) divergent self-energy contributions,
i.e. in lattice perturbation theory,
\begin{eqnarray}
\<L\>\simeq e^{-c_1N_\tau g^2+{O}(g^4)},
\label{div2}
\end{eqnarray}
where $1/aT=N_\tau$ and $c_1$ denotes the coefficient\cite{Karsch} of the
expansion in the bare coupling $g^2$. From this expression it follows, that
taking the continuum limit of (\ref{div2}) at fixed temperature leads to a
vanishing Polyakov loop expectation value even in the deconfined phase: The
divergence appearing on a lattice has to be removed by
renormalization\cite{Forcrand}.

Why is a renormalization of $\<L\>$ of interest? Firstly, $\<L\>$ acts as an
order parameter for the confinement-deconfinement phase transition which should
have a physical meaning. Secondly, a renormalized Polyakov loop can be used to
construct effective actions for the hadron dynamics near the phase
transition\cite{Pisarski}. And
thirdly, Polyakov loop correlation functions refer to free energies
containing the potential energy and entropy which are of interest in heavy
quark physics. For example investigations of meson and hadron properties via
the Polyakov loop and the Polyakov loop
correlation functions on a lattice show the need for both, the
renormalized Polyakov
loop  and the renormalized $n$-point Polyakov loop correlation functions.
\vspace{-0.cm}
\section{Towards renormalized $n$-point Polyakov loop correlations}
Recently we suggested\cite{Kacze} that a renormalization of the Polyakov loop
can be obtained through the renormalization of the quark-anti-quark free energy
calculated at short quark anti-quark distances. At short distances the finite
temperature
quark-anti-quark free energy is given by the zero temperature heavy-quark
potential. From this
property it follows that
the divergent self energy contribution in the
quark-anti-quark free energy at finite temperature can be removed through a
matching of its short distance behaviour to that of the heavy-quark potential at
zero temperature, which may be calculated perturbatively\cite{Necco}. Once
having done so, also the large distance behaviour of the
finite temperature free energy is fixed and can be used to extract the
renormalized Polyakov loop as no additional divergences are introduced
at finite temperature. Clearly, to each renormalization
prescription corresponds a renormalization constant (which in general depends
on the number of colors and
flavors) and it is worth noting that the Polyakov loop and the Polyakov loop
correlation functions are composite
operators and as such need their own renormalization constants and
conditions.

It turns out, however, that once having fixed the renormalization
constant for the Polyakov loop operator,
the renormalization scheme applies to any $n$-point Polyakov loop correlation
function at finite temperature. This
important feature is based on the (hierarchic) divergence structure of
$n$-point
Polyakov loop correlation functions which allows to remove the divergences
through one single renormalization constant. It thus follows that it is
sufficient to calculate only $2$-point Polyakov loop correlation functions at
finite temperature in order to extract the renormalized Polyakov loop and
the renormalization constant.

In the renormalization scheme described above, however, the renormalized
quantities are
defined only up to an arbitrary constant ($c$) in the zero temperature
potential. In terms of 2-point correlation
functions we may fix the zero temperature potential relatively
to the potential of the string picture at large quark
anti-quark
separations, for instance we define $c$ through
$V_{(c)}(R\gg1)\simeq\sigma R+c+{
  O}({1}/{R})$, where $V_{(c)}(R)$ denotes the zero
temperature potential and $\sigma$ is the string tension. Usually one takes the
standard Cornell potential ($c\equiv0$).

In order to be definite, we define the renormalized Polyakov
loop $L_{(c)}^R$ and its renormalization constant $Z_{(c)}^R(g^2)$ through
\begin{eqnarray}
L_{(c)}^{R}({\bf x})&\equiv&\prod\limits_{x_0=1}^{N_\tau}Z_{(c)}^R(g^2)U(x_0,{\bf
  x})\;=\;\left(Z_{(c)}^R(g^2)\right)^{N_\tau}L({\bf x}).
\label{Lrenformal}
\end{eqnarray}
A multiplicative introduction of the renormalization constant in the lattice
operator respect the center symmetry of $SU(N)$ and ensures that also the
expectation value of the renormalized Polyakov loop can be used as an
order parameter for the confinement-deconfinement phase transition.
Moreover, (\ref{Lrenformal}) implies
\begin{eqnarray}
&&\<\tilde{\mbox{Tr}}L^{R}_1({\bf
  x}_1)\;...\;\tilde{\mbox{Tr}}L^{R}_{nq}({\bf
  x}_{nq})\;...\;\tilde{\mbox{Tr}}L^{R \dagger}_{{\tilde n}\bar{q}}({\bf
  y}_{{\tilde n}\bar{q}})
\>\;=\;\left(Z_{(c)}^R(g^2)\right)^{(nq+{\tilde n}\bar{q})N_\tau}\<\; ...\;
  \>\nonumber\\
&&\qquad\qquad\qquad=\;\left(Z_{(c)}^R(g^2)\right)^{(nq+{\tilde
  n}\bar{q})N_\tau}e^{-\beta(F_{nq,
  {\tilde n}\bar{q}}-F_0)}.
\label{npoint}
\end{eqnarray}
The behaviour of (\ref{npoint}) at infinite quark separations
can be used to define the renormalized Polyakov loop, for instance in terms of
$2$-point Polyakov loop correlation functions,
\begin{eqnarray}
\frac{F^\infty_{(c)}(T)}{T}\equiv\lim_{R\to\infty}\frac{F_{(c)q\bar{q}}^R(R,T)}{T}
=-\ln|\<\left(Z_{(c)}^R(g^2)\right)^{N_\tau}\!\!\!L\>|^2=-\ln|\<L_{(c)}^{R}\>|^2.
\label{Lren}
\end{eqnarray}
We should note that (\ref{Lren}) respects the usual
colour structure of a quark anti-quark pair: It can form a colour singlet or
colour octet state. However, the renormalized
quark-anti-quark free
energies coincide\cite{Kacze,Kacze2} in all colour channels at infinite quark separation.
Consequently, the renormalized Polyakov loop expectation value, which can
be estimated from
$\exp(-F^\infty_{(c)}/2T)$, does not reflect a colour structure and its
 magnitude is
related to the free energy of a single heavy quark placed into a
thermal gluonic heat bath.
\section{Discussion of lattice results in $SU(N_c\equiv3)$}
\begin{figure}[t]
\centerline{\epsfxsize=7.8cm\epsfysize=6.9cm\epsfbox{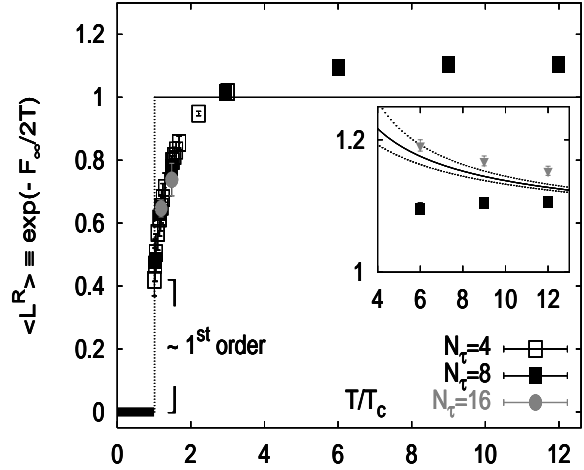}}
\vspace{-0.5cm}
\caption{The renormalized Polyakov loop expectation value as defined in
  (\ref{Lren}) calculated on lattices of size $32^3\times4,8,16$. The matching
  to the zero temperature potential utilizes the standard Cornell form
  ($c=0$). The inserted
  figure shows a comparison of
  $\<L^R\>$ for the case $c=0$ (squares) and $c=-T_c$ (triangles) at $T\gsim
  6T_c$ to (continuum) 1-loop perturbation theory.}
\label{free_energy}
\end{figure}
Lattice results\cite{Kacze} of the renormalized Polyakov loop are shown in
Fig.~1. The renormalized Polyakov loop expectation value acts as
an order parameter for the confinement-deconfinement phase transition and is
well behaved in the continuum limit. At high temperatures, $\<L^R\>$ approaches
values larger than unity, which is supposed to be its high temperature
limit. However, the renormalized Polyakov loop depends
(multiplicatively) on the fixing of the zero-T potential. In order to
demonstrate this dependence and to study contact with perturbation
theory\cite{Polyakov} we have inserted data with different
fixing constants, $c=0$ and $c=-T_c$ (see inserted figure).
Although the renormalization
scheme depends on this fixing, the renormalized Polyakov loop expectation value
approaches an 'universal' value at large temperatures: Indeed, for
different $c$ and $\tilde{c}$ it follows $\lim_{T\to\infty}\<L^R_{(\tilde
  c)}\>=\lim_{T\to\infty}\<L^R_{(c)}\>$.
Since a shift of the zero temperature potential corresponds to a
non-perturbative contribution in the renormalized Polyakov loop $<L^R\>$
we expect the renormalized Polyakov loop to coincide with (continuum)
perturbation theory at high temperatures where the shift does not influence the
magnitude of $\<L^R\>$. We therefore also expect
that the renormalized Polyakov loop defined in our scheme approaches unity at
infinite temperature.

\begin{figure}[t]
\begin{center}
\small
\begin{tabular}{||r||c|c|c|c|c|c|}\hline
$\beta\equiv6/g^2$    &4.100&4.127&4.154&4.179&4.200&4.229\\ \hline
$N_\tau=4$        &1.367(2)&1.368(3)&1.367(3)&1.368(2)&1.368(2)&1.369(2) \\ \hline\hline
$\beta\equiv6/g^2$    &4.321&4.343&4.365&4.400&4.600&4.839\\ \hline
$N_\tau=4$        &1.367(2)&1.366(1)&1.365(1)&1.364(1)&1.349(1)&1.329(1) \\ \hline\hline
$\beta\equiv6/g^2$    &4.5592&4.6605&4.8393&5.4261&6.0434&6.6450  \\ \hline
$N_\tau=8$ &1.355(1) &1.346(1) &1.330(1) &1.283(4) &1.242(2) &1.209(1)     \\ \hline
\end{tabular}
\end{center}
\caption{The non-perturbatively determined renormalization constant defined
  through $Z_{(0)}^R=(\<L^R_{(0)}/\<L\>)^{1/N_\tau}$.}
\label{table}
\end{figure}
The renormalization constant for the Polyakov
loop can be estimated in terms of $(\<L^R\>/\<L\>)^{1/N_\tau}$.
We only note here, that the values of $Z^R$ (see Fig.~2) are independent of
the lattice extent in time direction and
compensate the divergence appearing in (\ref{div2}). For small couplings $g^2$,
$\beta\equiv6/g^2\in[4.8,6.6]$, the data for $Z^R$ can be well described with a function
$f(g^2)=\exp(Q_2(8/3)g^2+Q_4g^4)$ with $Q_2=0.106$ and
$Q_4=-0.065$. These values are of the same magnitude as they appear in
lattice perturbation theory\cite{Karsch}.

\section{Conclusion}
We have shown that the non-perturbative renormalization scheme applies to
$n$-point Polyakov loop correlation functions once having fixed the
renormalization constant for one of these correlation functions. This can,
for instance, be done for the simplest
Polyakov loop correlation function, the $2$-point function which describes the
quark-anti-quark free energy. The
renormalization of the free energies is equivalent to having fixed the corresponding
partition function from which other physical quantities like the
potential energy\cite{Kacze2} can now be deduced.

\end{document}